# Non-destructive Three-dimensional Imaging of Artificially Degraded CdS Paints by Pump-probe Microscopy

Yue Zhou[1], David Grass[1], Warren S. Warren[1,2,3,4], and Martin C. Fischer*[1,2]

[1] Department of Chemistry, Duke University, Durham, NC 27708 USA
[2] Department of Physics, Duke University, Durham, NC 27708 USA
[3] Department of Biomedical Engineering, Duke University, Durham, NC 27708 USA
[4] Department of Radiology, Duke University, Durham, NC 27708 USA

**Abstract:** Cadmium sulfide (CdS) pigments have degraded in several well-known paintings, but the mechanisms of degradation have yet to be fully understood. Traditional non-destructive analysis techniques primarily focus on macroscopic degradation, whereas microscopic information is typically obtained with invasive techniques that require sample removal. Here, we demonstrate the use of pump-probe microscopy to nondestructively visualize the three-dimensional structure and degradation progress of CdS pigments in oil paints. CdS pigments, reproduced following historical synthesis methods, were artificially aged by exposure to high relative humidity (RH) and ultraviolet (UV) light. Pump-probe microscopy was applied to track the degradation progress in single grains, and volumetric imaging revealed early CdS degradation of small particles and on the surface of large particles. This indicates that the particle dimension influences the extent and evolution of degradation of historical CdS. In addition, the pump-probe signal decrease in degraded CdS is observable before visible changes to the eye, demonstrating that pump-probe microscopy is a promising tool to detect early-stage degradation in artworks. The observed degradation by pump-probe microscopy occurred through the conversion from CdS into $CdSO_4 \cdot xH_2O$, verified by both FTIR (Fourier-transform infrared) and XPS (X-ray photoelectron spectroscopy) experiment

## Introduction

Cadmium sulfide (CdS)-based yellow pigments, also known as cadmium yellow, are a group of important inorganic pigments in art history.[1] The introduction of these pigments to artists' palettes was followed by the improvement of industrial manufacturing in the 19th-20th century.[2] These pigments were favored by prominent artists, including Claude Monet,[3] Vincent van Gogh,[4] Edvard Munch,[5] Henri Matisse,[6] and Pablo Picasso,[7] due to their vivid colors and bright hues. However, many masterpieces by these artists have been found to suffer from CdS degradation such as fading, darkening, chalking, and flaking. The quality of historical CdS,[7] the presence of synthetic residues,[5] additives used by artists,[4] and environmental preservation conditions[8] can influence the degradation behaviors, complicating the conservation of artworks.

Great efforts have been made by art conservators and researchers to understand the mechanisms underlying the deterioration. CdS is a semiconductor with a direct bandgap of 2.42 eV (512 nm) (energy diagram shown in Figure SI2).[9] Within the bandgap, deep trap states originate from CdS defects, such as surface sulfur and cadmium vacancies.[10] These trap states are suspected to drive CdS degradation, as reported in a recent study of Pablo Picasso's Femme.[7] This effect was speculated to be more pronounced for small CdS pigments due to the increased surface-to-volume ratio.[7] In addition, photocatalytic activities of CdS nanoparticles with surrounding organic binders were suggested to be enhanced by an elevated number of defects, but further evidence is needed to fully understand the role of CdS particle size in the overall degradation process.[7]

The synthesis method can affect the quality of Cd-based pigments and the stability of paints. Historical CdS pigments were mainly produced by either a dry or a wet method.[1] The dry method involved the calcination of metallic cadmium, cadmium oxide, or cadmium carbonate with a stoichiometric amount or an excess of sulfur at 300-500°C; the resulting CdS was washed and ground prior to applications in paintings. The wet process precipitated CdS from sulfide (hydrogen sulfide, sodium sulfide, or barium sulfide) and cadmium salt (cadmium chloride, cadmium nitrate, or cadmium sulfate) solutions; the precipitate was then washed and used without further thermal treatment. The pigments produced by wet methods are more prone to degradation due to their poor crystalline structures, smaller dimensions, and the presence of synthetic residues and/or byproducts from insufficient washing procedures.[11] For example, cadmium chloride ($CdCl_2$) was detected in some extensively deteriorated sections in historical paintings.[6a-c, 12] However, the exact role of residues remains unclear, and additional studies are required to reveal their influences on CdS degradation.

Environmental conditions strongly affect the degradation of CdS paints. In artificial aging experiments, humidity was identified to be the primary cause of CdS degradation, while light and elevated temperatures could exacerbate degradation,[8] consistent with the hypothesis that soluble impurities affect degradation mechanisms.[5]

Traditional nondestructive techniques such as Raman, UV-Vis-NIR reflectance, X-ray fluorescence, and photoluminescence spectroscopy have been employed to study CdS degradation. However, due to their limitation in resolution, these techniques typically acquire only macroscale information. Synchrotron radiation-based X-ray spectro-microscopic methods[5] do provide high chemical and spatial resolution but are mostly restricted to analyzing cross-sectional samples.

Here, we demonstrate the use of pump-probe microscopy, a nonlinear optical technique, to nondestructively generate three-dimensional, high-resolution maps of paint structures and to track the degradation process on a microscopic scale. Traditionally used in biological imaging,[13] this technique has recently been applied in cultural heritage to create virtual cross-sections of paintings,[14] differentiate red organic dyes,[15] and identify the vermilion degradation product in a 14th-century painting.[16]

The principle of pump-probe microscopy can be found in a previous review[17] and will only be briefly described here. The experimental setup is shown in Figure 1a. The pump-probe approach generates contrast by nonlinear optical interactions between the sample and two synchronized femtosecond pulse trains, named pump and probe. The pump pulse train is amplitude-modulated and superimposed with the probe pulse train before entering the laser scanning microscope. Pump and probe pulse trains interact nonlinearly with the sample in the focal region, thereby transferring some amplitude modulation from the pump to the probe pulse train (see Figure 1b). The backscattered probe light is collected while the pump light is removed with a filter, and the amplitude modulation of the probe pulse is measured with a photodiode and a lock-in amplifier. Changing the inter-pulse time delay ($\tau$) between pump and probe results in characteristic transient absorption curves. In this study, we operate the microscope at a 720 nm pump and 817 nm probe wavelength combination with 0.4 mW of optical power in each pulse train. Many nonlinear interactions are accessible with pump-probe microscopy (see Figure SI3), but for our wavelength combination the predominant interaction between laser pulses and CdS is two-photon absorption (TPA, see Figure 1c). TPA involves a virtual energy state, and the sample absorbs one photon from each pulse only when pump and probe pulses arrive simultaneously at the sample ($\tau$ = 0 ps). The multiphoton nature of pump-probe microscopy allows for virtual sectioning and volumetric imaging in heterogeneous paint samples. By scanning the focus of the laser beams across the sample in both lateral directions and moving the sample in the axial direction, we can non-destructively map CdS grains within paints. This allows us to monitor CdS changes during artificial aging by recording the reduction of TPA signals.

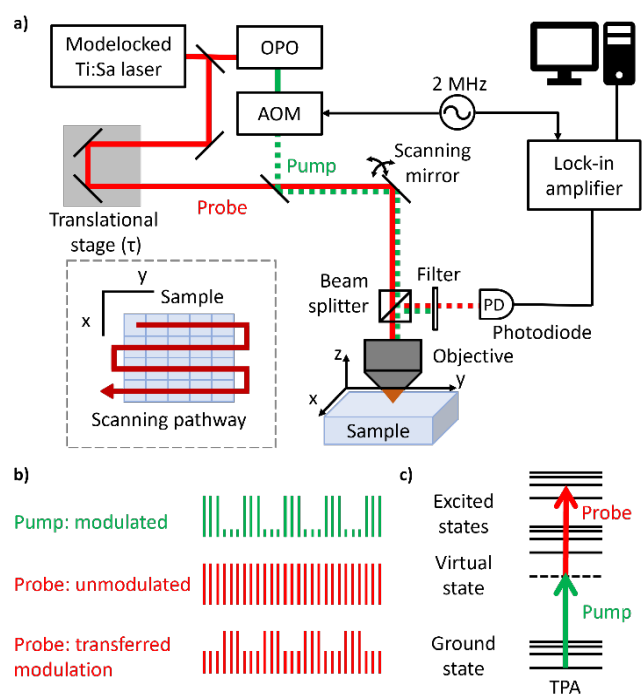

**Figure 1.** Pump-probe microscopy. a) Simplified schematic of the experimental setup. For details, see the Supporting Information. b) Modulation transfer scheme. c) Two-photon absorption (TPA). The dashed line indicates a virtual state.

In this study, we present pump-probe imaging of artificially aged CdS paints which are prepared according to historical recipes. We demonstrate that pump-probe microscopy can noninvasively monitor the degradation inside single grains during artificial aging with micrometer-scale resolution. In addition, early-stage alterations on CdS mock-up paints can be recognized before visual changes are apparent, highlighting the potential of pump-probe microscopy to detect the degradation of seemingly well-preserved artworks.

## Results and Discussion

### Reproduction and Characterization of Historical CdS Pigments

To simulate the degradation in historical paintings through artificial aging, CdS was reproduced following historical wet methods. CdS pigments synthesized from wet methods are more susceptible to degradation because of the poor crystalline structure and potential synthetic residues.[5, 11a] $CdCl_2$ was chosen as a starting agent because chloride was found in degraded regions in multiple paintings,[5] making it one of the most plausible ingredient of the historical manufacturing process. $Na_2S$ was selected as sulfur source since previous research showed that CdS made from $Na_2S$ had similar morphology and photoluminescence properties as historical pigments.[18] This wet method reaction is described below. More details about the synthetic procedure can be found in the Supporting Information.

$$CdCl_2 + Na_2S \rightarrow CdS + 2NaCl$$

CdS (Sigma Aldrich) was purchased as reference sample, and both synthesized and commercial CdS powders were first characterized and then prepared as oil paints to be artificially degraded. Figure 2 displays commercially available (top row) and synthesized (bottom row) CdS samples: powders (a,d), derived linseed oil paints (b,e), and microscopic images (c,f). The particle morphology and crystal forms were characterized by SEM-EDS (scanning electron microscopy) images (Figure 3) and X-ray diffraction (XRD) spectra (Figure 4).

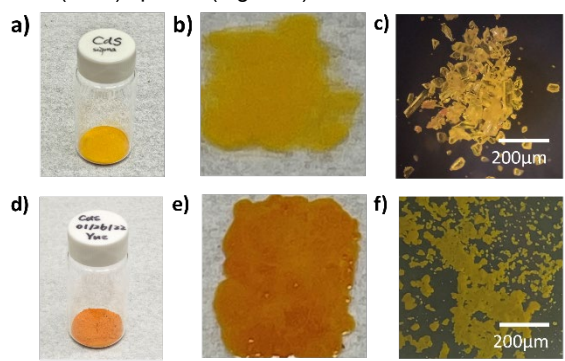

**Figure 2.** Commercial and synthesized CdS. Commercial CdS (Sigma Aldrich): a) powder, b) linseed oil paint, and c) microscopic image. Synthesized CdS: d)

powder, e) linseed oil paint, and f) microscopic image. Commercial CdS contains large, semitransparent crystals, while synthesized CdS displays smaller grains without clear crystalline structures.

The SEM image of Figure 3a indicates that commercial CdS has a smooth surface with a well-defined crystalline structure in agreement with the microscopic photos. The EDS (energy disperse X-ray spectroscopy) spectrum of commercial CdS in Figure 3c shows a high purity of CdS with a minor amount of C (which may originate from sample processing). The synthesized CdS is composed of grains with a rough surface and nanometer-sized particles, as shown in Figure 3b. The corresponding EDS spectrum shows small contributions from C, O, Si (possibly comes from the imaging substrate), and negligible amount of Na (synthetic residue) in addition to the expected Cd and S. There was no detectable amount of Cl, which suggests an effective washing process in synthesis procedures. The magnified SEM image (shown in Figure SI3) reveals the existence of aggregated grains from nano-scale particles. The nano-dimensional and aggregated particles have a larger surface-to-volume ratio than commercial CdS grains, which could reduce the specular reflectance and lead to a darker color than commercial CdS.

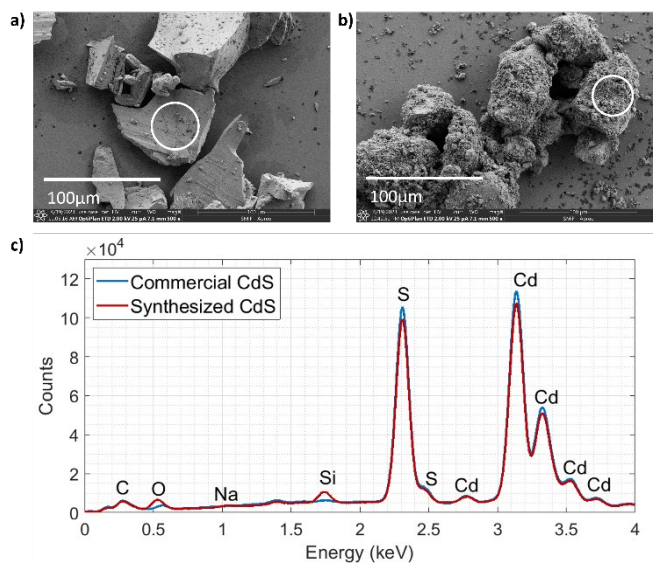

**Figure 3.** SEM-EDS analysis of commercial and synthesized CdS. SEM images of a) commercial and b) synthesized CdS. c) EDS spectra of commercial (blue) and synthesized (red) CdS. Both show the elemental components in the circled regions of their corresponding SEM images.

Figure 4a shows XRD spectra of the two CdS powders with reference lines[19] for hexagonal and cubic CdS. The commercial CdS shows a clear hexagonal structure, while the synthesized CdS exhibits a more cubic and amorphous structure. The crystal structure difference can also contribute to the color difference of two CdS (where commercial CdS shows a lighter color).[20] In addition, the broad XRD peaks of synthesized CdS are a sign of imperfect crystal lattice and nanometric crystal size.[21] UV-Vis-NIR reflectance spectra shown in Figure SI5a also confirm the color difference between two CdS samples. The sigmoidal-shaped reflectance spectrum of synthesized CdS is another sign of poor crystal structure.[5, 22] The fluorescence spectra (shown in Figure SI5b) show that commercial CdS has a sharp band gap emission at 512 nm which is typical for bulk CdS. The synthesized CdS presents no band gap emission, similar to archived historical CdS pigments examined in a previous study.[2]

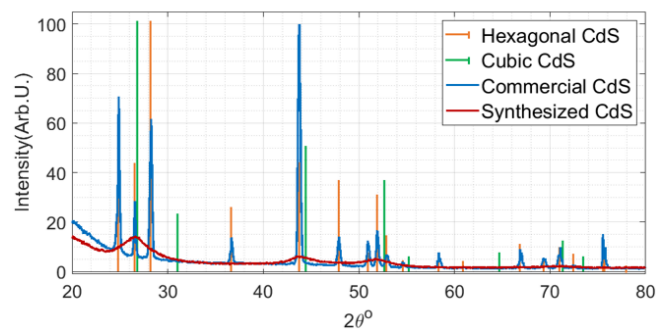

**Figure 4.** XRD spectra of CdS and references (hexagonal and cubic) crystal structures. Commercial CdS shows strong hexagonal peaks, while synthesized CdS shows a cubic and amorphous structure.

## Artificial Aging of CdS Paints

Artificial aging of both CdS paints was performed in a homemade aging chamber.[8] The CdS paints were applied on microscope slides, and aged by UV light (~400 nm) irradiation and exposure to high relative humidity (RH) levels (~75%) for four weeks. These samples were labeled as sample$_{UV+RH}$. Control experiments were also performed in which identical samples were exposed to only UV light (sample$_{UV}$) or only high RH levels (sample$_{RH}$) or left in a dark and dry environment (sample$_{N/A}$). Pump-probe microscopy was utilized to monitor the change in CdS paints during the four weeks of aging. The commercial CdS samples exposed to both UV light and high RH levels (sample$^{com}$) showed only negligible signs of degradation; therefore, we focus on the synthesized CdS sample for the following sections, unless otherwise noted. More details on sample preparation, aging chamber design, and aging process are documented in the Supporting Information.

Table 1 presents photos of synthesized CdS paints before and after aging with light and high RH levels. The paints did not exhibit apparent signs of degradation during the first two weeks, but slight alterations, such as a global loss of surface oil gloss, became visible after four weeks. The samples exposed to only high RH levels retained their oil gloss, suggesting a lower degree of degradation in comparison to the UV-exposed sample.

**Table 1.** Photographs of synthesized CdS oil paints before and after aging[a].

| Aging conditions | Unaged sample | Aged for one week | Aged for two weeks | Aged for four weeks |
|---|---|---|---|---|
| UV + high RH level | | | | |
| High RH level | | | | |

[a] Sample$_{UV}$ and sample$_{N/A}$ are not shown here since they did not show visible signs of degradation over four weeks. This is consistent with pump-probe experiments discussed later.

The UV-Vis-NIR reflectance spectra shown in Figure 5 agree with visual observations. The reflectance curve of sample$_{UV+RH}$ exhibited a blue shift indicating a universal color change of the paint to lighter shades of yellow.

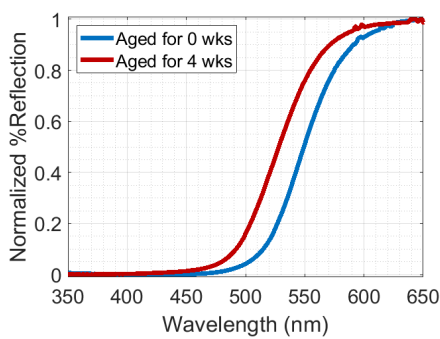

**Figure 5.** UV-Vis-NIR reflectance spectra of sample$_{UV+RH}$ before and after aging.

## Pump-probe Signatures of CdS Paints and Visualization of CdS Degradation

To monitor the degradation progress on a microscopic scale, pump-probe microscopy was used to image CdS paints after various durations of artificial aging. Pump-probe images of the same region of interest (ROI) on sample$_{UV+RH}$ before, after two weeks, and after four weeks of aging are shown in Figures 6a, b, and c. CdS grains are dominated by an instantaneous TPA signal, which can be seen in the transient absorption curves averaged over the rectangular areas plotted in Figure 6d. The signal decreases by 20%-30% during the first two weeks of aging and by more than 80% during weeks two to four.

To map the change in pump-probe signals of CdS during aging, we co-registered pump-probe images of aged and unaged paints and computed the ratio of TPA signal between aged and unaged CdS. Figure 6e shows a pixel-by-pixel ratio image between two weeks of aging and no aging (Figures 6b divided by 6a). This ratio image presents the signal loss over two weeks of artificial aging. The false coloring is chosen such that unaltered regions are displayed in blue, and areas with drastically diminished signals are displayed in red. Green shows a moderate level of degradation. Similarly, Figure 6f shows the ratio image between four weeks of aging and two weeks of aging (Figure 6c divided by 6b). Beyond four weeks of aging, the pump-probe signal decreased to unobservable levels, limited by the signal-to-noise performance of our microscope. The pump-probe signatures of degradation are more severe in the second two weeks of aging (weeks two to four). Changes after two weeks were not yet observable by visual inspection or by bright field microscopy, but were obvious in the pump-probe data, emphasizing the ability of pump-probe to detect early-stage degradation.

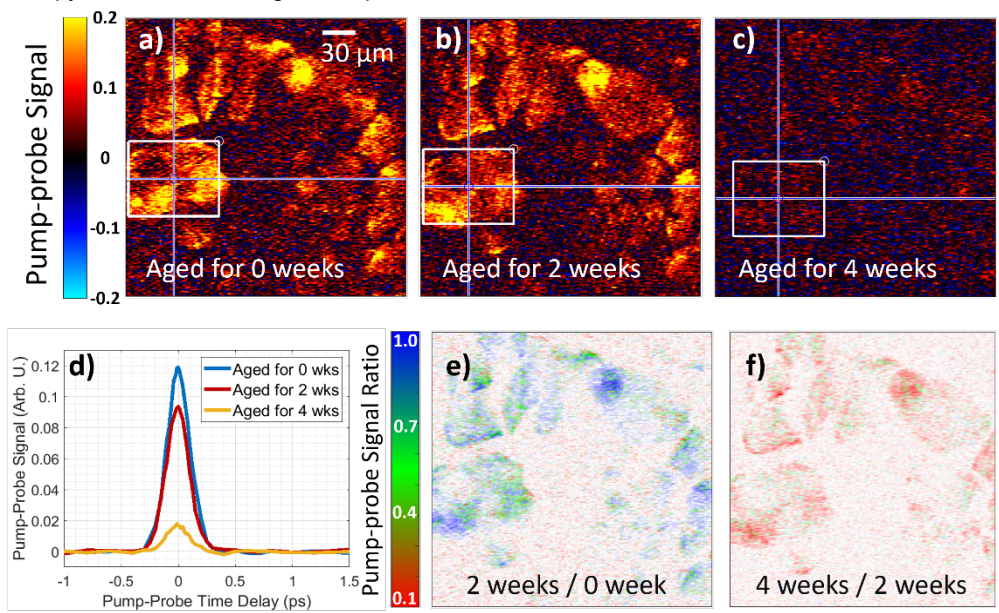

**Figure 6.** Pump-probe false-color images of sample$_{UV+RH}$ a) before aging, b) aged for 2 weeks, and c) aged for 4 weeks d) Pump-probe transient absorption curves spatially averaged over the white rectangle in the pump-probe images. Pump-probe pixel-by-pixel ratio images of e) 2 weeks over 0 weeks of aging and f) 4 weeks over 2 weeks of aging.

To visualize early-stage degradation in three dimensions on the surface and the interior of the paint, the sample$_{UV+RH}$ was investigated after aging for one week (no degradation visible in color or gloss, neither by eye nor by bright field microscopy). Figure 7 shows a volume ratio image between one week of aging and no aging. A video of a full three-dimensional scan through this volume can be found in the Supporting Information. Small grains and the surface of large grains exhibited a moderate degree of CdS degradation shown in green. In addition, the inside of large grains stays unaffected, shown in blue. These observations indicate a particle size influence on the degradation of CdS: smaller particles experience more degradation in a shorter period of time. This result supports the previous hypothesis that CdS paints containing nano-sized particles experience more intense degradation.[7]

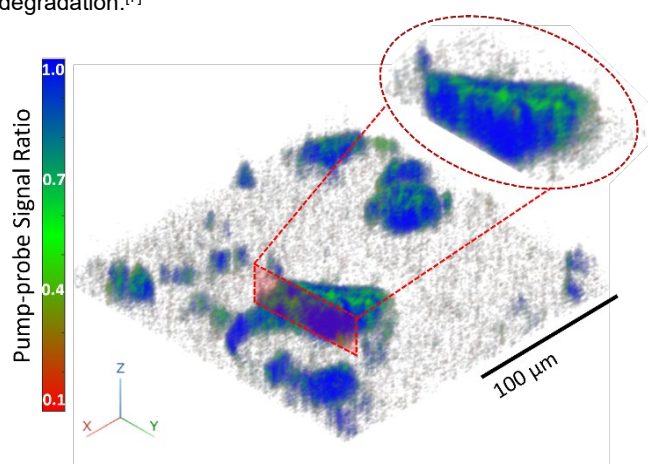

**Figure 7.** Three-dimensional ratio image of sample$_{UV+RH}$ aged for one week. The small grains and the surface of large grains experienced moderate levels of degradation, while the inside of large grains showed no signs of degradation. The insert shows a cross-section of a large CdS grain that displays no degradation (blue) inside with moderate degradation (green) on the surface.

To understand the individual effects of moisture and light on the degradation of synthesized CdS pigments, Figure 8 shows the loss of signals in the control samples after four weeks of aging. Sample$_{RH}$ exhibited moderate degradation with 30%-60% pump-probe signal decline (Figure 8a), which is less compared with the 80% signal decrease of sample$_{UV+RH}$. This result further supports previous findings that the moisture mainly triggers the degradation of CdS paints.[5, 8] Sample$_{UV}$ showed only minor signs of degradation after four weeks (Figure 8b), indicating that light itself is not a sufficient factor to degrade CdS within the duration of exposure. Sample$_{N/A}$ showed no sign of pump-probe signal decrease within four weeks (Figure 8c), indicating that the pump-probe signal change did not result from factors other than light and moisture.

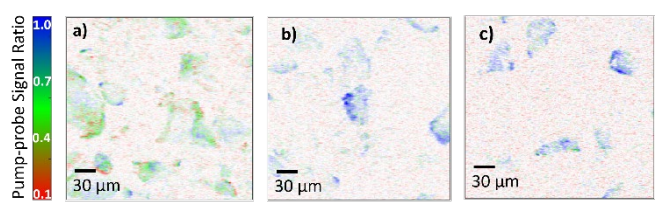

**Figure 8.** Pump-probe ratio images between samples being aged for four weeks and before aging. a) Only exposed to moisture (sample$_{RH}$). b) Only exposed to UV light (sample$_{UV}$). c) No exposure (kept in dark and low RH levels, sample$_{N/A}$).

The commercial CdS paint (sample$^{com}$) showed no detectable pump-probe signal decrease even after eight weeks of aging by both light and high RH levels (see Figure 9), demonstrating the stability of modern CdS samples with well-defined crystal structure and negligible amounts of defects.

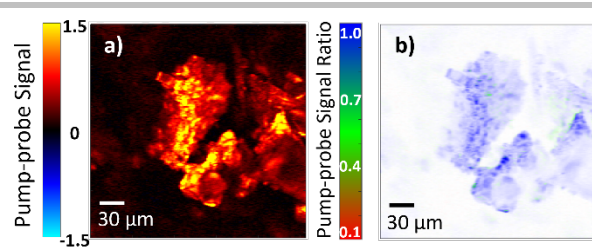

**Figure 9.** False-color a) pump-probe image and b) pixel-by-pixel ratio image (aged for 8 weeks / before aging) of commercial CdS (sample$^{com}$).

## FTIR and XPS Techniques Confirmed the Degradation of CdS Paints

FTIR spectroscopy is usually regarded as a nondestructive technique. However, the attenuated total reflectance (ATR) mode requires physical contact between the sample and detection crystal, which would have distorted the paint layer (rearranged individual pigments in our paint samples) and altered the surfaces. Therefore, FTIR-ATR experiments were only performed after the 4-week aging period and after pump-probe imaging. The common degradation products and impurities found in historical CdS paintings, including $CdSO_4$, $CdCO_3$, and $CdCl_2$,[4-5, 12] were also studied individually to serve as reference chemicals and compare with aged paints. The product details and FTIR spectra (Figure SI6) of reference chemicals can be found in the Supporting Information.

The FTIR spectra of sample$_{N/A}$ feature notable CH stretching (at 2853 and 2924 cm$^{-1}$) and CO stretching peaks (1737 cm$^{-1}$). These peaks are typical fingerprints of linseed oil binder used to prepare paint samples. For sample$_{RH}$, the oil peaks moderately declined and a new peak of free fatty acids at 1708-1713 cm$^{-1}$ appeared, revealing the breakdown of the binder; no other changes were observed from the FTIR spectra of this sample. Sample$_{UV+RH}$ displayed a more distinguished decrease of linseed oil peaks and an increase of free fatty acids, representing a higher degree of oil degradation. In addition, new spectral peaks appeared at 1173, 1063, 992, 882, 822, 653, 618, 606, 589, 520, and 481 cm$^{-1}$ (see Figure 10b), which match the commercial $CdSO_4 \cdot xH_2O$ reference. Therefore, we conclude that the CdS was converted into $CdSO_4 \cdot xH_2O$ under light exposure and high RH levels. No other reference chemicals or degradation products from the literature[8] were identified in degraded paints. FTIR analysis unambiguously confirms that we successfully degraded CdS. These spectra also demonstrate that the transition from CdS to $CdSO_4$ happens predominantly under exposure to both light and high RH levels.

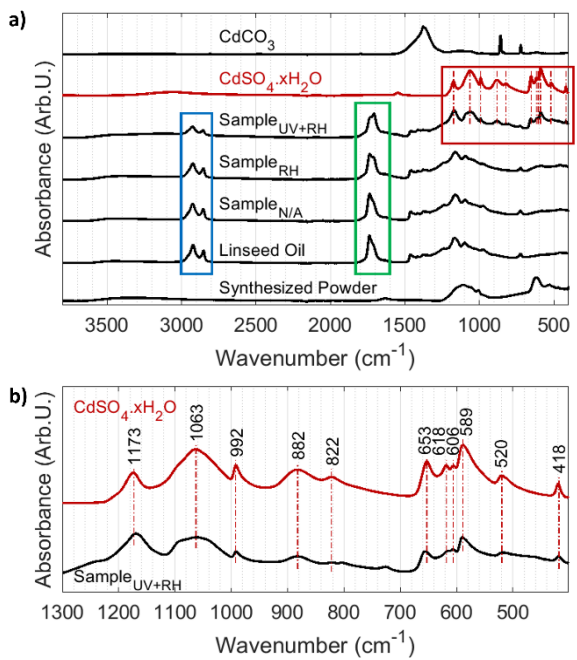

**Figure 10.** FTIR spectra of synthesized CdS powder, linseed oil binder, unaged and aged synthesized CdS paints, $CdSO_4 \cdot xH_2O$ (Sigma Aldrich), and $CdCO_3$ (Sigma Aldrich). a) Blue rectangle: CH stretching peaks at 2853 $cm^{-1}$ and 2924 $cm^{-1}$ decreased after aging, indicating the degradation of linseed oil binder; Green rectangle: decrease of CO stretching peak (1737 $cm^{-1}$) and increase of free fatty acids peaks (1708-1713 $cm^{-1}$); Red rectangle: aged CdS paint shows pronounced overlap with $CdSO_4 \cdot xH_2O$ (details in 10b). b) Sample$_{UV+RH}$ shows pronounced correspondence with $CdSO_4 \cdot xH_2O$, indicating the degradation products of the paint.

In addition, commercial CdS paint (sample$^{com}$) exhibits no sign of degradation other than the increase of free fatty acid peaks (Figure SI7). This further supports the pump-probe findings that commercial CdS did not degrade during aging.

XPS experiments were carried out on synthesized CdS pigment powders and sample$_{UV+RH}$ to compare the chemical state of sulfur components. The paint sample before aging was not investigated in this experiment since x-rays get absorbed in the oil-rich surface of newly prepared paints. The S 2p spectra can reflect the amount of each sulfur species in two samples, thus indicating the evolution of sulfur components. The spectra show 87% of $S^{2-}$ and 13% of $SO_4^{2-}$ in synthesized CdS powders (Figure 11a). After artificial aging, the $S^{2-}$ on the sample surface decreased to 33%, and the $SO_4^{2-}$ increased to 67% (Figure 11b), emphasizing the oxidation of sulfur from $S^{2-}$ to $SO_4^{2-}$ during aging. Again, the XPS spectra verified the conclusions drawn from pump-probe microscopy and FTIR experiments, that the CdS turned into $CdSO_4$ during aging by UV light and high RH levels. This conclusion is also consistent with the findings in previous research.[8]

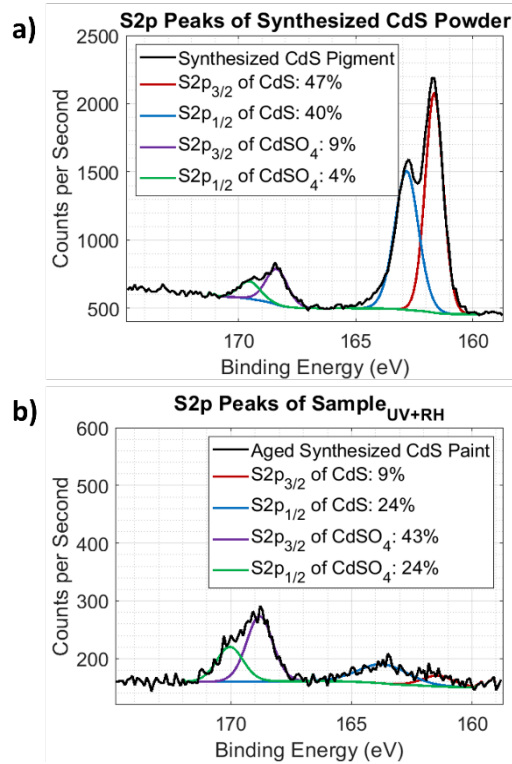

**Figure 11.** XPS S2p spectra of a) synthesized CdS powders and b) sample$_{UV+RH}$. The red, blue, purple, and green curves are fitted data of S2p peaks of CdS and $CdSO_4$ components. The molar/area percentage values are labeled for each component. The composition of $CdSO_4$ increased from 13% to 67% after artificial aging.

## Conclusion

We demonstrated the use of pump-probe microscopy to visualize CdS degradation in cadmium yellow paints, in particular the onset of degradation before visible changes can be observed. CdS were produced following historical synthesis methods and exhibits distinctive properties: poor crystalline structures, aggregated grains with nano-dimensional particles, and no band gap emission. These properties are similar to those of archived historical pigments making them ideal candidates for artificial aging experiments. Mock-up paints prepared from synthesized CdS were artificially aged to generate degradation similar to those found in historical artworks. We showed that pump-probe microscopy can nondestructively create 3-dimensional visualizations of CdS degradation in paints on the microscopic scale. In summary, we draw the following conclusions:

1. Pump-probe microscopy can detect the onset of CdS degradation through a decline of the TPA signal. This signal decrease is already observable after a few weeks of aging, suggesting that degradation happens microscopically before macroscopic observations are possible.
2. In-situ, nondestructive pump-probe imaging can track degradation of paint samples during the aging process. Three-dimensional pump-probe images show that degradation starts primarily in smaller grains and on the surface of large grains.

3. The stability of CdS paints is affected by the degree of crystallinity of pigments and aging conditions. Commercial CdS with well-defined crystal structures displayed no change in pump-probe signal, even after eight weeks of aging, while synthesized CdS with poor crystal structures showed obvious changes in pump-probe signatures within one to two weeks. For synthesized CdS paints, humidity is necessary to induce degradation, while light is an accelerator to turn CdS into $CdSO_4 \cdot xH_2O$.

Overall, pump-probe microscopy can provide in-situ, early-stage detection of CdS degradation on the micrometer scale. We believe that pump-probe microscopy is a valuable technique to nondestructively investigate pigment degradation, for example to study the role of leftover synthetic residues or secondary products (such as $CdCl_2$ and $CdSO_4$) in the degradation process of CdS paints. Future investigations can be carried out on the microscopic degradation behaviors of paint samples with impurities or mixed with other historical pigments.

## Supporting Information

The authors have cited additional references within the Supporting Information.[8-9, 17a, 23]

## Acknowledgements

This material is based upon work supported by the National Science Foundation Division of Chemistry under Award No. CHE-xxxxxxx (M.C.F.). We would like to thank Dr. John Delaney, Dr. Mathieu Thoury, Dr. Daniela Comelli, and Dr. Marta Ghirardello for their valuable assistance and discussions. We also thank Heidi Kastenholz and Dr. Jake Lindale for helpful discussions.

The SEM-EDS, XRD, and XPS experiments were performed at the Duke University Shared Materials Instrumentation Facility (SMIF), a member of the North Carolina Research Triangle Nanotechnology Network (RTNN), which is supported by the National Science Foundation (award number ECCS-xxxxxxx) as part of the National Nanotechnology Coordinated Infrastructure (NNCI).

**Keywords:** Cadmium sulfide • Degradation • Pump-probe microscopy • 3-dimensional visualization • Noninvasive in-situ imaging

## Reference

[1] I. Fiedler and M. Bayard in *Cadmium Yellows, Oranges and Reds*, Vol. 1 (Ed. R. L. Feller), National Gallery of Art, Washington, **1986**, pp. 65–108.
[2] M. Ghirardello, S. Mosca, J. Marti-Rujas, L. Nardo, A. Burnstock, A. Nevin, M. Bondani, L. Toniolo, G. Valentini and D. Comelli, *Analytical Chemistry* **2018**, *90*, 10771-10779.
[3] A. Roy, *National Gallery Technical Bulletin* **2007**, *27*, 58-68.
[4] G. Van der Snickt, K. Janssens, J. Dik, W. De Nolf, F. Vanmeert, J. Jaroszewicz, M. Cotte, G. Falkenberg and L. Van der Loeff, *Analytical Chemistry* **2012**, *84*, 10221-10228.
[5] L. Monico, L. Cartechini, F. Rosi, A. Chieli, C. Grazia, S. De Meyer, G. Nuyts, F. Vanmeert, K. Janssens, M. Cotte, W. De Nolf, G. Falkenberg, I. C. A. Sandu, E. S. Tveit, J. Mass, R. P. de Freitas, A. Romani and C. Miliani, *Science Advances* **2020**, *6*, eaay3514.
[6] a) J. L. Mass, R. Opila, B. Buckley, M. Cotte, J. Church and A. Mehta, *Applied Physics A* **2013**, *111*, 59-68; b) J. Mass, J. Sedlmair, C. S. Patterson, D. Carson, B. Buckley and C. Hirschmugl, *Analyst* **2013**, *138*, 6032-6043; c) E. Pouyet, M. Cotte, B. Fayard, M. Salomé, F. Meirer, A. Mehta, E. S. Uffelman, A. Hull, F. Vanmeert, J. Kieffer, M. Burghammer, K. Janssens, F. Sette and J. Mass, *Applied Physics A* **2015**, *121*, 967-980; d) Z. E. Voras, K. deGhetaldi, M. B. Wiggins, B. Buckley, B. Baade, J. L. Mass and T. P. Beebe, *Applied Physics A* **2015**, *121*, 1015-1030.
[7] D. Comelli, D. MacLennan, M. Ghirardello, A. Phenix, C. Schmidt Patterson, H. Khanjian, M. Gross, G. Valentini, K. Trentelman and A. Nevin, *Analytical Chemistry* **2019**, *91*, 3421-3428.
[8] L. Monico, A. Chieli, S. De Meyer, M. Cotte, W. de Nolf, G. Falkenberg, K. Janssens, A. Romani and C. Miliani, *Chemistry – A European Journal* **2018**, *24*, 11584-11593.
[9] M. Thoury, J. K. Delaney, E. R. de la Rie, M. Palmer, K. Morales and J. Krueger, *Applied Spectroscopy* **2011**, *65*, 939-951.
[10] A. Veamatahau, B. Jiang, T. Seifert, S. Makuta, K. Latham, M. Kanehara, T. Teranishi and Y. Tachibana, *Physical Chemistry Chemical Physics* **2015**, *17*, 2850-2858.
[11] a) A. H. Church, *The Chemistry of Paints and Painting*, Seeley, Service & Co. Limited, **1915**, p; b) A. P. Laurie, *Facts about Processes, Pigments and Vehicles: A Manual for Art Student*, Macmillan and Co., London, **1895**, p.
[12] G. Van der Snickt, J. Dik, M. Cotte, K. Janssens, J. Jaroszewicz, W. De Nolf, J. Groenewegen and L. Van der Loeff, *Analytical Chemistry* **2009**, *81*, 2600-2610.
[13] a) J. Kuk-Youn, D. Simone, C. F. Martin, C. Z. Kevin, J. Xiaomeng, Y. Jin and S. W. Warren, *Journal of Biomedical Optics* **2019**, *24*, 051414; b) D. Grass, G. M. Beasley, M. C. Fischer, M. A. Selim, Y. Zhou and W. S. Warren, *Optics Express* **2022**, *30*, 31852-31862.
[14] T. E. Villafana, W. P. Brown, J. K. Delaney, M. Palmer, W. S. Warren and M. C. Fischer, *Proceedings of the National Academy of Sciences* **2014**, *111*, 1708-1713.
[15] J. Yu, W. S. Warren and M. C. Fischer, *Analytical Chemistry* **2018**, *90*, 12686-12691.
[16] J. Yu, W. S. Warren and M. C. Fischer, *Science Advances* **2019**, *5*, eaaw3136.
[17] a) M. C. Fischer, J. W. Wilson, F. E. Robles and W. S. Warren, *Rev Sci Instrum* **2016**, *87*, 031101; b) M. J. Melo, M. Ghirardello, A. Candeo, D. Comelli, C. Manzoni, M. Thoury, M. Réfrégiers, M. Oujja, M. Sanz, A. D. Fovo, R. Fontana, M. Castillejo, Y. Zhou, H. Kastenholz, W. S. Warren and M. C. Fischer in *Laboratory Instrumentation*, Springer International Publishing, Cham, pp. 1-36.
[18] M. Ghirardello in *Study of the Physical-chemical Properties of Cadmium Yellow: Understanding How Synthesis Methods Impact on Paint Stability*, Vol. Doctoral Degree The Polytechnic University of Milan, Milan, Italy, **2021**, p. 161.
[19] a) H. Sowa, *Solid State Sciences* **2005**, *7*, 73-78; b) N. A. Noor, N. Ikram, S. Ali, S. Nazir, S. M. Alay-e-Abbas and A. Shaukat, *Journal of Alloys and Compounds* **2010**, *507*, 356-363.
[20] M. Ghirardello, V. Otero, D. Comelli, L. Toniolo, D. Dellasega, L. Nessi, M. Cantoni, G. Valentini, A. Nevin and M. J. Melo, *Dyes and Pigments* **2021**, *186*, 108998.
[21] P. B. Raja, K. R. Munusamy, V. Perumal and M. N. M. Ibrahim in *5 - Characterization of nanomaterial used in nanobioremediation*, Eds.: H. M. N. Iqbal, M. Bilal and T. A. Nguyen), Elsevier, **2022**, pp. 57-83.
[22] H. Deng and J. M. Hossenlopp, *The Journal of Physical Chemistry B* **2005**, *109*, 66-73.
[23] J. Jiang, W. S. Warren and M. C. Fischer, *Optics Express* **2020**, *28*, 11259-11266.

# Supporting Information

## Experimental Procedures

### 1.1 Materials

Commercial cadmium sulfide (CdS) was purchased from Sigma Aldrich. Two chemicals were used for CdS synthesis: sodium sulfide nonahydrate ($Na_2S \cdot 9H_2O$) from Sigma Aldrich and cadmium chloride ($CdCl_2$) from Fluka Analytical. Four commercial chemicals from Sigma Aldrich were used as the reference for FTIR experiments: cadmium sulfate ($CdSO_4$), cadmium sulfate hydrate ($CdSO_4 \cdot xH_2O$), cadmium chloride hydrate ($CdCl_2 \cdot xH_2O$), and cadmium carbonate ($CdCO_3$).

### 1.2 Synthesis of reproduced CdS following historical recipes

The CdS was prepared by adding a 50ml 0.05M $Na_2S$ solution to a 50ml 0.05M $CdCl_2$ solution under stirring. The mixture was settled for 2 hours and filtered with paper. The precipitate was washed with water and ethanol. The powder was dried in a dark and cool environment for 2 weeks.

### 1.3 Preparation of CdS oil paints

The CdS mock-up paints were prepared by mixing the commercial or synthesized CdS powders with linseed oil (Williamsburg Oil) in a 2:1 mass ratio. The paints were applied to microscope slides and dried for 2 months in a dark and cool environment.

### 1.4 Artificial aging of CdS oil paints

The artificial aging experiments were carried out in a home-made aging chamber[8] equipped with a fluorescent light bar and a $KNO_3$ saturated solution. The scheme of the aging chamber is shown in Figure SI1a. The chamber consists of a sealed container, and the mock-ups were exposed to UVA-Vis light (see Figure SI1b for the spectrum), and/or high relative humidity (RH) levels (>75%), at room temperature (27 °C). The irradiance of the sample position was 1 mW cm$^{-2}$. The total aging time was four to eight weeks.

The control experiments were performed in the same chamber with different conditions. CdS paints were exposed to the following conditions: light and high RH levels, light and low RH levels (<45%), only high RH levels, kept in dark (by covering samples with a black aluminum foil), and low RH levels.

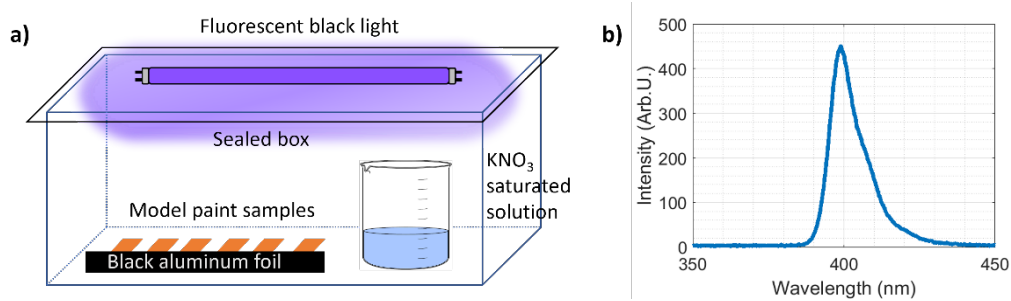

**Figure SI1.** Scheme of the a) CdS paint aging chamber and b) the spectrum of aging UV-Vis light.

### 1.5 Analytical methods

The paint samples were analyzed before aging, after two weeks, after four weeks, and after eight weeks of aging with pump-probe microscopy, and after every four weeks with UV-Vis spectrometry (with a diffuse reflectance accessory). Finaly, the aged paints were analyzed with FTIR (Fourier-transform infrared) and XPS (X-ray photoelectron) spectroscopy.

#### 1.5.1 Pump-probe microscopy

Two ultrashort (150 fs pulse length) laser pulses of distinct colors, the pump and the probe, are generated from an 80 MHz mode-locked Ti: sapphire laser (Chameleon, Coherent) and an optical parametric oscillator (Mira-OPO, Coherent). The pump / probe wavelength combination was tuned at 720 / 817 nm at 0.4 mW for each pulse. The pump pulse train was amplitude-modulated at 2 MHz by an acousto-optic modulator (AOM), while the probe remains unmodulated. A translational stage tunes the path length of one pulse train and thus controls the inter-pulse delay (τ). These two pulse trains are aligned collinearly, focused on the sample with an objective lens (Olympus UPlanApo 20x objective, NA = 0.7), and scanned through the sample areas. At the focal point, the modulation is transferred from the pump to the probe by nonlinear optical interactions. The transmitted or back-scattered probe light can be gathered by a photodiode, demodulated by a lock-in amplifier, and acquired by a computer as a function of the surface position (x, y), the depth (z), and the inter-pulse delay (τ). The focal size of the beams are around 0.6 µm.

### 1.5.2 Scanning electron microscopy with energy dispersive x-ray spectroscopy

A ThermoFisher Scientific Apreo S with EDS detector was used for morphology (primary voltage; 2 keV; working distance: 7 mm) and elemental (primary voltage: 20 keV; working distance: 10 mm) characterization of CdS. Pigments were dispersed in ethanol and a drop of the solution was placed onto a silicon substrate to dry in the air before the experiments.

### 1.5.3 X-ray diffraction analysis

X-ray diffractograms were acquired with a PANalytical X'Pert Pro high-resolution XRD system with a 1/2° slit, a step size of 0.05°, an acquisition time per step of 1 s, and Cu-Kα radiation (generator voltage: 45 kV; tube current: 40 mA).

### 1.5.4 UV-Vis-NIR reflectance Spectroscopy

A Cary 5000 UV-Vis-NIR spectrometer with diffuse reflectance accessory was applied to obtain linear reflectance spectra. The spectra were collected from 350 to 800 nm with a step size of 1 nm. A white Spectralon disc covered with the same microscope slide was employed for the reference measurement.

### 1.5.5 Steady-state photoluminescence analysis

The photoluminescence spectra were obtained by a Zeiss 710 Scanning Laser Microscope with multispectral mode. The measurements were performed with an 817nm laser source (Chameleon, Coherent). The detection range was 421-723 nm with a 9.7 nm step size. Microscope slides with pure pigments attached were studied.

### 1.5.6 X-ray photoelectron spectroscopy (XPS)

The XPS measurements were performed by a Kratos Analytical Axis Ultra system. The x-ray source was Al-Kα, and the pass energy was 20 eV with a step size of 0.1 eV. The measurements were focused on the S2p core level to obtain the stoichiometry of the CdS samples before and after aging. The experimental data were fitted by the CasaXPS software with a convolution of Gaussian line shapes.

### 1.5.7 ATR-FTIR spectroscopy

The FTIR spectra were acquired by a Thermo Scientific™ Nicolet™ iS50 FTIR Spectrometer in ATR mode. The aged paint samples are attached to the built-in diamond ATR and recorded in the 4000-400 cm-1 range at 0.482 cm-1 resolution.

## 2. Results and Discussion

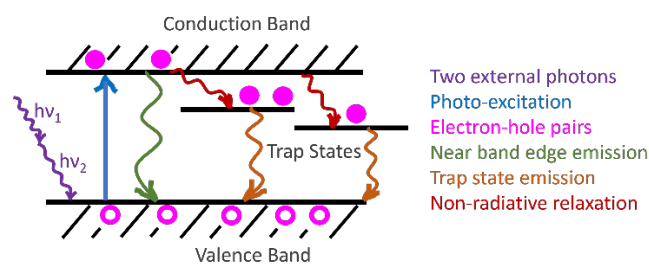

**Figure SI2.** Energy diagram of bulk CdS semiconductor. External photons with energies higher than the bandgap (here two-photon excitation is shown in the figure) energy can excite electrons to the conduction band, leaving holes in the valence band. Excited electrons relax to the valence band either by near-band edge emissions or by non-radiative decay to two deep trap states in the bandgap followed by trap state emission. Near band edge emissions occur on the picosecond timescale, while emissions from trap states occur on the order of microseconds.[9]

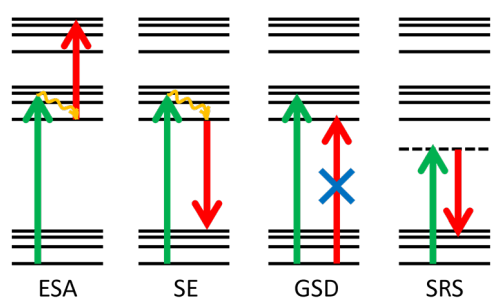

**Figure SI3.** Some nonlinear optical interactions accessible with pump-probe microscopy. Excited-state absorption (ESA; absorption of the probe photon after the pump photon), Stimulated Raman scattering (SRS, the gain or loss of the probe), stimulated emission (SE; emission of a probe photon after the absorption of a pump photon), and ground state depletion (GSD; the competition between absorptions of the pump and probe photons). Solid lines and dashed lines indicate discrete energy levels and virtual states, respectively.[17a]

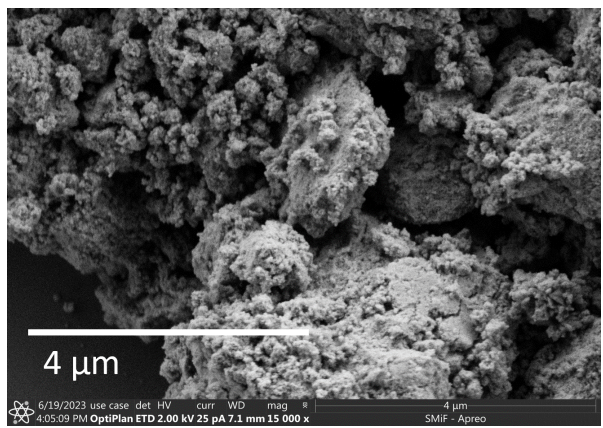

**Figure SI4.** SEM image of synthesized CdS magnified by 15000 times. The CdS grains are aggregated from smaller, nano-sized particles.

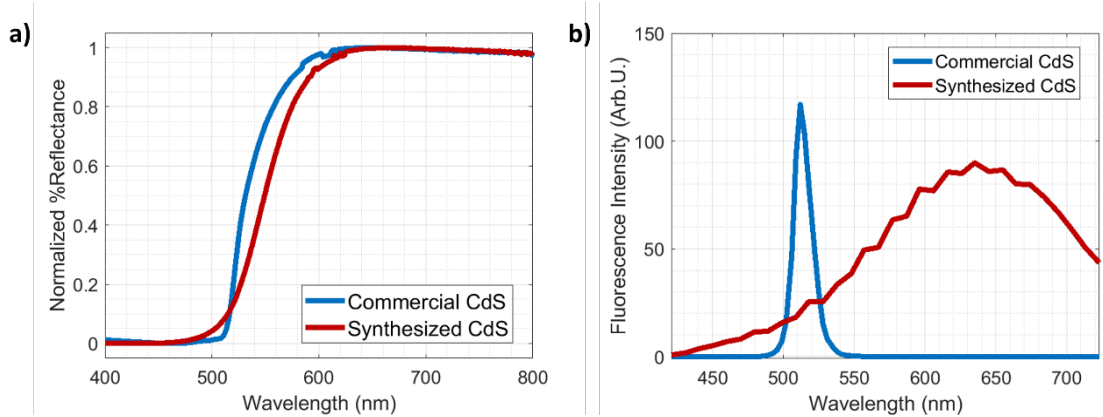

**Figure SI5.** a) UV-Vis reflectance spectra and b) fluorescence emission spectra of commercial and synthesized CdS. The sigmoidal-shaped reflectance spectrum of synthesized CdS is a sign of poor crystal structure. Commercial CdS shows typical sharp band gap emission of bulk CdS, while the synthesized CdS exhibited a broad "red luminescence" peak centered around 635nm.

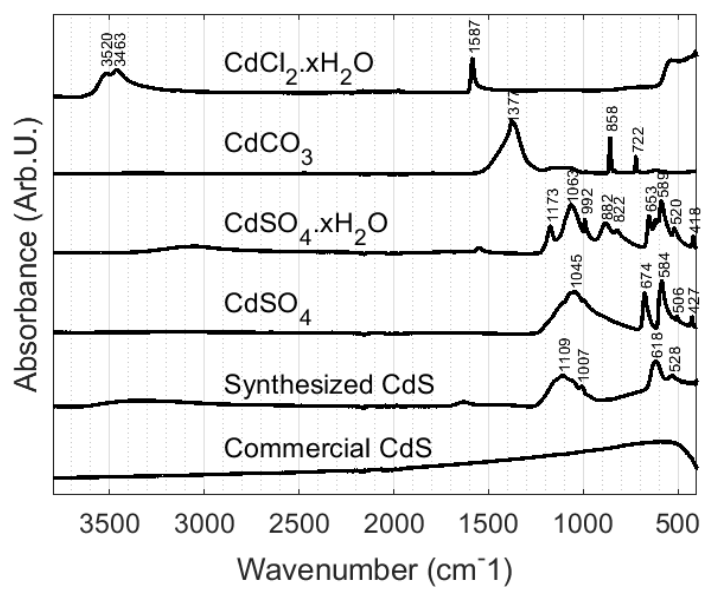

**Figure SI6.** FTIR spectra of reference chemicals (CdCl$_2$·xH$_2$O, CdCO$_3$, CdSO$_4$, CdSO$_4$·xH$_2$O) and two CdS powders.

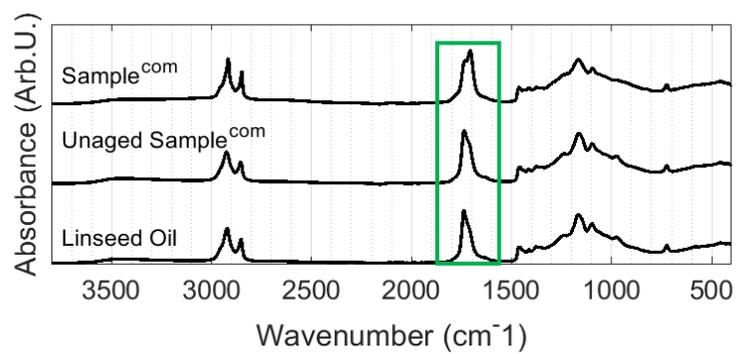

**Figure SI7.** FTIR spectra of commercial CdS after aging for eight weeks. The green rectangle highlights the decrease of CO stretching peak (1737 cm$^{-1}$) and increase of free fatty acids peaks (1708-1713 cm$^{-1}$), suggesting the change of linseed oil. No other peak appears, which indicates negligible degradation of commercial CdS.